\documentclass[smallextended]{my-svjour3}

\usepackage{booktabs} 

\usepackage{xcolor}
\usepackage{siunitx}
\usepackage{multirow}
\usepackage{xspace}
\usepackage{balance}
\usepackage[font=rm]{caption}
\usepackage{algorithm}
\usepackage[noend]{algpseudocode}
\usepackage{verbatim}
\usepackage{graphicx}
\usepackage{tabularx}
\usepackage{enumitem}
\usepackage[round,semicolon,sort]{natbib}
\usepackage{shortvrb}
\usepackage{url}
\usepackage{tikz}
\usetikzlibrary{matrix,decorations.pathreplacing,calc,positioning}

\setlist[itemize,1]{leftmargin=5mm,itemsep=0mm}
\setlist[enumerate,1]{leftmargin=5mm,itemsep=0mm}

\newcommand{\var}[1]{\mbox{\emph{#1}}}

\newcommand{\myparagraph}[1]{\subsection{#1}}
\newcommand{\mycaption}[1]{\caption{{\rm{#1}}}}

\usepackage{xcolor}

\newcommand{\Num}[1]{\mbox{$N(#1)$}}
\newcommand{\Lab}[1]{\mbox{\sf{#1}}}
\newcommand{\Labpair}[2]{\langle\Lab{#1},\Lab{#2}\rangle}
\newcommand{\company}{AmaBaiBinGoo}
\newcommand{\RB}{\mbox{\emph{RB}}}

\newcommand{\degrees}[1]{\mbox{$#1{}^{\circ}$}}

\journalname{Copyright is held by the author}

\begin{document}
\title{Batch Evaluation Metrics in Information Retrieval:\\
Measures, Scales, and Meaning}
\titlerunning{Batch Evaluation Metrics in IR: Measures, Scales, and Meaning}

\author{Alistair Moffat}

\institute{Alistair Moffat \at
            School of Computing and Information Systems,
	    The University of Melbourne, Melbourne, Australia \\
            \email{ammoffat@unimelb.edu.au}\\
	    ORCiD: 0000-0002-6638-0232
}

\date{7 July 2022}

\maketitle

\begin{abstract}
A sequence of recent papers has considered the role of measurement
scales in information retrieval (IR) experimentation, and presented
the argument that (only) uniform-step interval scales should be used,
and hence that well-known metrics such as reciprocal rank, expected
reciprocal rank, normalized discounted cumulative gain, and average
precision, should be either discarded as measurement tools, or
adapted so that their metric values lie at uniformly-spaced points on
the number line.
These papers paint a rather bleak picture of past decades of IR
evaluation, at odds with the community's overall emphasis on
practical experimentation and measurable improvement.

Our purpose in this work is to challenge that position.
In particular, we argue that mappings from categorical and ordinal
data to sets of points on the number line are valid provided there is
an external reason for each target point to have been selected.
We first consider the general role of measurement scales, and of
categorical, ordinal, interval, ratio, and absolute data collections.
In connection with the first two of those categories we also provide
examples of the knowledge that is captured and represented by numeric
mappings to the real number line.
Focusing then on information retrieval, we argue that document
rankings are categorical data, and that the role of an effectiveness
metric is to provide a single value that represents the usefulness to
a user or population of users of any given ranking, with usefulness
able to be represented as a continuous variable on a ratio scale.
That is, we argue that current IR metrics are well-founded, and,
moreover, that those metrics are more meaningful in their current
form than in the proposed ``intervalized'' versions.
\end{abstract}

\section{Introduction}
\label{sec-intro}

We use {\emph{measurement}} to capture data about some attribute or
observation of a real-world phenomenon.
For example, a thermometer measures temperature so that we have
guidance as to how hot or cold we might feel if we went outdoors; and
weather forecasts and climate summaries predict temperatures based on
datasets of past observations so that we can make informed choices
about our future activities.
All other things being equal, a beach-side holiday at a location with
an average daytime temperature of {\degrees{25}C} is likely to be
preferable to one at a location with an average daytime temperature of
{\degrees{10}C}.

If a measurement is to be useful, it should be connected to the
attribute it refers to, and allow inferences to be derived from sets
of measurements that are both predictive of, and connected to, the
reality that is being measured.
The extent to which inferences can be derived from measurements is,
in part, determined by the {\emph{scale}} that is employed, defined
as the way in which observed attributes are represented by measured
surrogates.
For example, patients in a hospital may be asked to rate their pain
on a scale of one to ten, with ``perceived patient pain'' the
attribute, and a set of ten ordered categories as the corresponding
measurement.
Pain amelioration medications might then in part be selected in
accordance with the measurement a patient reports, even though
patients are not calibrated ``pain-o-meters''.

{\citet{stevens46}} described four {\emph{scales of measurement}},
and enumerated their {\emph{permissible operations}} in terms of what
can be legitimately concluded about the behavior of the underlying
attribute, given knowledge of the measurement, or of a set of
measurements.
In the case of a ``ten point'' pain assessment, for example, if
yesterday a patient said ``eight'' and today they say ``four'', we
can be confident that their perceived pain has decreased, but should
not say that it has halved -- the conclusion ``has decreased'' is
permissible, but the conclusion ``has halved'' is not, and nor is the
conclusion ``has gone down by four''.
In the case of ordinal measures (which is what the ten point pain
scale is), comparison of measurements is permitted as a valid
reflection of the relative values of the corresponding underlying
attributes, but the taking of ratios and differences is not (at
least, not necessarily).
Indeed, a patient's subjective pain scale might be highly non-linear.
Section~\ref{sec-background} provides more detail of
Stevens' hierarchy of scales of measurement.

In a sequence of recent papers,
{\citet{ffp19ieeekade,ffl20irj,fff21ieeeaccess}} have explored the
way in which experimental measurement is carried out in information
retrieval (IR), with particular emphasis on the use of
{\emph{interval scales}}, one of Stevens' four scale categories.
In a somewhat bleak assessment, {\citeauthor{fff21ieeeaccess}}
conclude that IR has, by and large, employed inappropriate
measurement scales for the last several decades, and (at risk of
putting words into their mouths) that the discipline now has an
opportunity to lift its game.
In their most recent contribution, {\citet{fff21ieeeaccess}} propose
that each existing IR metric be ``intervalized'' in order to be
rendered sound, with raw metric values mapped to evenly distributed
points on the number line, the goal being to create a corresponding
adjusted metric on equi-interval measurement scale.

Our goal here is to express a more optimistic view of IR evaluation
and IR measurement, and to challenge the conclusions of
{\citeauthor{fff21ieeeaccess}}
We argue that the majority of existing IR metrics are in fact
well-founded in terms of measurement and, for the most part, achieve
what it was that they were designed to capture.
We also differ from {\citeauthor{fff21ieeeaccess}} in regard to the
worth of ``intervalization'' as a corrective mechanism.

In Section~\ref{sec-background} Stevens' measurement scale typology
is explained, and examples of categorical, ordinal, and interval
scales are provided.
Section~\ref{sec-serps} then focuses in IR evaluation, and explains
how IR measures are defined; assesses them against Stevens'
hierarchy; summarizes {\citeauthor{fff21ieeeaccess}}'s proposals; and
explains why we diverge from them.
Section~\ref{sec-other} then considers other related work, and other
aspects of IR experimental methodology that researchers and
practitioners alike should bear in mind.

\section{Background}
\label{sec-background}

The distinction between {\emph{categorical}}, {\emph{ordinal}},
{\emph{interval}}, and {\emph{ratio}} measures was introduced by
{\citet{stevens46}} and is now widely summarized in textbooks and
online resources.\footnote{See, for example,
{\url{https://en.wikipedia.org/wiki/Level_of_measurement}}.}
This section provides an overview of those four classes, starting
with categorical and ordinal measures, and examples thereof; then
taking a diversion into a parable in which professors' salaries are
tabulated; and finishing with a description of interval and ratio
measures.
Section~\ref{sec-serps} then considers the ways in which measures are
applied to evaluation in information retrieval.

\myparagraph{Categorical Data}

\begin{table}[t]
\centering
\mycaption{Two sets of categorical (or nominal) labels:
(a) the standard set of country codes, used perhaps to record country
of birth in personnel data; and
(b) a set of academic work categories at a university.
\label{tbl-categorical}}
\begin{tabular}{cl}

\toprule
Label
	& Class
\\
\midrule
\Lab{AU} & Australia
\\
\Lab{CL} & Chile
\\
\Lab{CN} & China
\\
\Lab{IT} & Italy
\\
\Lab{JP} & Japan
\\
\bottomrule
\\[-2.0ex]
\multicolumn{2}{c}{(a) Countries of birth}
\\[2ex]

\toprule
Label
	& Class
\\
\midrule
\Lab{AM} & administrative and management
\\
\Lab{RO} & research focused
\\
\Lab{TO} & teaching focused
\\
\Lab{TR} & teaching and research
\\
\bottomrule
\\[-2.0ex]
\multicolumn{2}{c}{(b) Academic employment types in a university}

\end{tabular}
 \end{table}

In a categorical (or nominal) measure each object in the data
collection of interest (a {\emph{dataset}}) is assigned to a single
class, with the classes identified via a set of {\emph{class
labels}}.
Table~\ref{tbl-categorical} provides two examples.
In Table~\ref{tbl-categorical}(a) the classes are {\emph{countries}},
perhaps reflecting birth locations of university employees, and the
class labels are two letter acronyms; and in
Table~\ref{tbl-categorical}(b) the classes are {\emph{academic work
categories}} in a university, and the labels are again two letter
acronyms.
For example, the birth countries of a group of ten professors at a
university might be represented by the dataset:
\[
	\{ \Lab{AU}, \Lab{CN}, \Lab{AU}, \Lab{IT}, \Lab{CL},
		\Lab{CL}, \Lab{CN}, \Lab{AU}, \Lab{IT}, \Lab{JP} \} \, .
\]
The cardinality of each class across the dataset can be tabulated and
used to summarize fractions (and hence allow representation as pie
charts, including pie charts in which the segments are ordered from
largest to smallest), and the mode (most frequent class) can be
reported, as can other class ranks based on frequency of occurrence.
For example, in the example dataset, the most common birth country is
{\Lab{AU}}, accounting for $3/10$ of the professors; similarly (in
the context of Table~\ref{tbl-categorical}(b)), a university might
report that ``research focused'' {\Lab{RO}} staff are its second
largest category, accounting for $31$\% of its employees.

In categorical data there is no meaningful ordering between the
classes, and the only operations that can be applied to data items
are equality and inequality testing ($=$ and $\not=$).
The classes themselves can be ordered by considering their labels to
be ASCII strings, as has been done in the two table sections, but
that is purely for presentational convenience, and not an intrinsic
feature of the data.
If the class labels changed (or even if they didn't) the table rows
could be reordered, without affecting the validity or accuracy of any
conclusions drawn from the dataset being discussed.

This absence of ordering between the classes means that the median
(and other percentiles), and arithmetic and geometric averages, are
meaningless concepts -- a fact immediately grasped when considering
the questions ``average country of birth'' and ``median work
category'' of the ten professors mentioned above.
Moreover, the inappropriateness of computing medians and means
remains even if the class labels ``look'' like numbers.
For example, suppose that the international phone dialing prefixes
were used as the class labels, rather than the two letter acronyms:
{\Lab{+61}} for Australia, {\Lab{+56}} for Chile, and so on (let's
ignore the fact that {\Lab{+1}} actually covers several countries).
The same dataset of ten professors would now be described as:
\[
	\{ \Lab{+61}, \Lab{+86}, \Lab{+61}, \Lab{+39}, \Lab{+56},
		\Lab{+56}, \Lab{+86}, \Lab{+61}, \Lab{+39}, \Lab{+81}
	\} \, ,
\]
but it is still {\emph{not}} permissible to compute the median or
mean.

When categorical data is combined into ordered tuples -- for example,
the pairing ``$\langle$country of birth, work category$\rangle$'' --
the set of tuples is also categorical data.

\begin{table}[t]
\centering
\mycaption{Four sets of ordinal labels, and for each
set, one possible mapping between the set's categories and the real
number line:
(a) the set of available options (radio buttons) in a survey about
frequency of alcohol consumption, with $\Num{\cdot}$ in ``risk
points'';
(b) the set of professorial ranks at some university, with
$\Num{\cdot}$ in ``salary, dollars per week'';
(c) the set of summative recommendations (radio buttons) available to
referees during a peer review process, with $\Num{\cdot}$ a numeric
mapping; and
(d) the set of relevance labels employed in a document judging
process as part of an IR evaluation, with $\Num{\cdot}$ a gain
expressed in units of ``utility''.
\label{tbl-ordinal}}
\begin{tabular}{clc}

\toprule
Label
	& Class
		& $\Num{\cdot}$
\\
\midrule
\Lab{A0}
	& never
		& $0$
\\
\Lab{A1}
	& monthly or less
		& $1$
\\
\Lab{A2}
	& two to four times a month
		& $2$
\\
\Lab{A3}
	& two to three times a week
		& $3$
\\
\Lab{A4}
	& four or more times per week
		& $4$
\\
\bottomrule
\\[-2.0ex]
\multicolumn{3}{c}{(a) Screening for alcohol consumption (one question of
several)}
\\[-0.7ex]
\multicolumn{3}{c}{\tiny{Source: https://www.uptodate.com/contents/calculator-alcohol-consumption-screening-}}
\\[-1.3ex]
\multicolumn{3}{c}{\tiny{audit-questionnaire-in-adults-patient-education}}
\\[2ex]

\toprule
Label
        & Class
		& $\Num{\cdot}$
\\
\midrule
\Lab{P1}
	& junior professor
		& $\$700$
\\
\Lab{P2}
	& assistant professor		
		& $\$750$
\\
\Lab{P3}
	& associate professor	
		& $\$850$
\\
\Lab{P4}
	& full professor
		& $\$1000$
\\
\bottomrule
\\[-2.0ex]
\multicolumn{3}{c}{(b) Professorial ranks}
\\[2ex]

\toprule
Label
        & Class
		& $\Num{\cdot}$
\\
\midrule
\Lab{SR}
	& strong reject
		& $-3$
\\
\Lab{R}
	& reject		
		& $-2$
\\
\Lab{WR}
	& weak reject	
		& $-1$
\\
\Lab{WA}
	& weak accept	
		& $+1$
\\
\Lab{A}
	& accept		
		& $+2$
\\
\Lab{SA}
	& strong accept	
		& $+3$
\\
\bottomrule
\\[-2.0ex]
\multicolumn{3}{c}{(c) Referee evaluations during peer review}
\\[2ex]

\toprule
Label
        & Class
		& $\Num{\cdot}$
\\
\midrule
\Lab{G0}
	& not relevant
		& $0.00$
\\
\Lab{G1}
	& somewhat relevant
		& $0.25$
\\
\Lab{G2}
	& relevant
		& $0.75$
\\
\Lab{G3}
	& highly relevant
		& $1.00$
\\
\bottomrule
\\[-2.0ex]
\multicolumn{3}{c}{(d) Grades for document relevance to a topic}
\\[1ex]

\end{tabular}
 \end{table}

\myparagraph{Ordinal Data}

Ordinal data classes result if strict inequality as well as equality
are permissible operators for comparing data items, that is, if all
of $=$, $\not=$, $<$, and~$>$ are operational.
Table~\ref{tbl-ordinal} gives four instances of ordinal class labels
and class descriptions; the final column headed $\Num{\cdot}$ will be
discussed shortly.
The first section of Table~\ref{tbl-ordinal} is taken from an online
``alcohol risk screening'' assessment, and is one of a suite of
questions that collectively ask about frequency (the one shown),
intensity (the amount of alcohol consumed at each session), and
impact (physical or emotional damage to self or to relationships with
family/friends).
There is a clear ordering, with category {\Lab{A0}} involving less
frequent alcohol consumption than categories {\Lab{A1}}, {\Lab{A2}},
and so on.

The fact that the classes are ordered means that cumulative
statistics are permissible.
For example, a university might report (in reference to
Table~\ref{tbl-ordinal}(b)) that $70$\% of its professors are at
level {\Lab{P2}} or higher.
The same university might also ask students to rate courses via a
five-point Likert scale, using the class labels ``strongly
disagree'', ``disagree'', ``neutral'', ``agree'', and ``strongly
agree'' in response to a statement ``this course was well taught''.
It would then be permissible to compute a dissatisfaction score for
each course by summing the percentage of ``strongly disagree'' and
``disagree'' responses, to focus pedagogical interventions on courses
where students are least satisfied.

The ordered classes mean that it is valid to identify the smallest
(\var{min}) and largest (\var{max}) item in any dataset, and also
permissible to sort a dataset into ``order''.
Medians of ordinal-scale datasets may also be calculated, albeit with
a degree of caution.
For example, in the dataset
$\{ \Lab{P1}, \Lab{P2}, \Lab{P3}, \Lab{P4}\}$
it is unclear what median value should be reported, but inferring it
to be ``$\Lab{P2.5}$'' via the ``arithmetic'' $(\Lab{P2}+\Lab{P3})/2$
over the two middle points in the sorted arrangement is clearly
absurd.
The best that can be said in this example is that the median lies in
the interval $[{\Lab{P2}}, {\Lab{P3}}]$.\footnote{In detail: a class
label $m$ is a median of a dataset $X$ if half (or more) of the
members of $X$ are $\le m$ and half (or more) of the members of $X$
are $\ge m$.
In the example that means that both $\Lab{P2}$ and $\Lab{P3}$ are
medians; and implies that in the dataset
$X=\{\Lab{P1},\Lab{P1},\Lab{P4},\Lab{P4}\}$ all four class labels
$\Lab{P1}$, $\Lab{P2}$, $\Lab{P3}$, and $\Lab{P4}$ are medians.
It also means that in the numeric dataset $X=\{1,1,4,4\}$
{\emph{every}} value $1\le m \le 4$ is a median.
The use of $2.5$ in this numeric case is then a convention that
isolates a single value amongst the infinite number of
possibilities.}

Ordinal data can be plotted as frequencies or cumulative frequencies
in a bar chart, with one bar per class, and the class labels on the
horizontal (domain) axis ordered according to the known $<$ ordering
between the classes.
Ordinal data can also be plotted as fractions within a pie chart,
since the operations that may be applied to categorical data are all
still available.
Moreover, in the case of ordinal data the pie segments can be placed
in a meaningful sequence based on the class ordering, or placed in
size order, or placed with no ordering at all.

In the absence of specific additional information, tuples based on
ordinal data must be regarded as being categorical data.
Suppose, for example, that a university asked an alcohol screening
question of its professors, to create a set of tuples ``$\langle$alcohol
consumption, professorial rank$\rangle$''.
We might feel justified in concluding that $\Labpair{A1}{P2}$ comes
``before'' the pair $\Labpair{A2}{P4}$, since both dimensions agree
on that ordering.
But we would have no ability to put $\Labpair{A3}{P3}$ and
$\Labpair{A2}{P4}$ into ``order''; therefore, the pairs must be
categories.
This important point will be returned to in Section~\ref{sec-serps}
when we consider IR evaluation.

\myparagraph{Numerical Transformations}

When the number of ordinal classes is small (for example,
radio-button surveys and Likert scales), the median is a relatively
blunt and non-discriminating tool; and an {\emph{ordinal to numeric
mapping}} is sometimes used to transform the class labels into
numbers that can be processed arithmetically (and perhaps
statistically).
For example, the five Likert dis/agreement class labels from
``strongly disagree'' to ``strongly agree'' might be converted to the
numeric values ``1'' to ``5'' so that ``average agreement'' can be
computed.

Similarly, each section in Table~\ref{tbl-ordinal} shows one such
possible mapping, denoted $\Num{\cdot}$.
For example, in Table~\ref{tbl-ordinal}(a) each answer (to this and
each other question in the survey instrument) is assigned a ``risk
points'' value, and a sum is computed over the answers across the set
of screening questions, to indicate the extent to which the survey
respondent is likely to be affected by alcohol-induced health and
social problems.
Similarly, in Table~\ref{tbl-ordinal}(c), the sum of the
$\Num{\cdot}$ values over the pool of referees assigned to each paper
might be computed, and used as an overall assessment, with negative
sums reflecting net ``rejection'', and positive sums indicating net
``acceptance''.
Despite the apparent ease with which such mappings can be
constructed, care is required, and each of those four example
mappings is just one instance of an infinite variety of possibilities
that could be devised and then argued for.

\myparagraph{Processing the Professorial Payroll} 

We now turn to a more detailed example involving an ordinal to
numeric mapping.
Suppose that, in the context of Table~\ref{tbl-ordinal}(b), some
university has a total of ten professors, with academic positions
given by the dataset:
\[
	\{  \Lab{P3},\Lab{P2},\Lab{P4},\Lab{P2},\Lab{P2},
	    \Lab{P2},\Lab{P1},\Lab{P3},\Lab{P4},\Lab{P2} \}\, ;
\]
and that the university wants to include that information into
its annual report via a small table:
\begin{center}
\begin{tabular}{l cccc}
\toprule

Class      & \Lab{P1} & \Lab{P2} & \Lab{P3} & \Lab{P4}
\\
Count      & 1 & 5 & 2 & 2
\\
\bottomrule
\end{tabular}
\end{center}
This is legitimate, since it is permitted to tabulate occurrence
frequencies in both categorical and ordinal datasets.
There is no sense of there being an ``average'' professor (although
many of our students would regard us as being ``mean''!);\ but the
median is a permissible statistic, and in this dataset there is no
ambiguity, the median is $\Lab{P2}$.
Suppose further that the next table of the annual report lists the
current weekly salaries for those four professorial ranks:
\begin{center}
\begin{tabular}{l cccc}
\toprule

Class      & \Lab{P1} & \Lab{P2} & \Lab{P3} & \Lab{P4}
\\
Salary      & \$700 & \$750 & \$850 & \$1000
\\
\bottomrule
\end{tabular}
\end{center}
This table is thus an ordinal to numeric mapping that allows
professorial ranks to be converted to numbers, and hence for the
dataset of ten professorial ranks to be mapped to a dataset of ten
weekly salaries:
\[
	\{  \$850, \$750, \$1000, \$750, \$750,
	    \$750, \$700, \$850, \$1000, \$750 \}\,.
\]
Calculation of the median salary over the professors is a legitimate
and correct operation; it is clearly $\$750$ per week.
The university could add that statistic to its annual report with a
clear conscience.

Now suppose that one of the {\Lab{P2}} professors is promoted to
level {\Lab{P3}} before the annual report is finalized.
What becomes of the median salary?
For a even-sized set of numbers the usual convention is to take the
mid-point between the two middle values (see, for example,
{\citet[Section 4.2]{hays1994}}); after the promotion, that
computation yields a median of $(\$750+\$850)/2=\$800$ per week, or
$\$50$ per week higher than it was previously.

Finally, the university also decides to include the total salary
being paid across the set of professors.
In a separate work area the annual report's editor prepares this
table, to reflect the situation after the successful promotion:
\begin{center}
\begin{tabular}{l cccc}
\toprule
Class      & \Lab{P1} & \Lab{P2} & \Lab{P3} & \Lab{P4}
\\
Count      & $1$ & $4$ & $3$ & $2$
\\
Salary      & \$700 & \$750 & \$850 & \$1000
\\
Payment      & \$700 & \$3000 & \$2550 & \$2000
\\
\bottomrule
\end{tabular}
\end{center}
The editor then sums the bottom row to get a total weekly salary cost
of $\$8250$, finalizes the annual report, and sends it to the
printer.

The very first copy arrives on the Provost's desk just a few days
later.
Worried about the budget, the Provost looks at these various
statistics, including the fact that there are ten professors and a
total salary cost of $\$8250$ per week, and concludes that the
average weekly salary (since even Provosts can divide by ten in their
heads) per professor is $\$825$.
It never crosses the Provost's mind to ponder the fact that the
original data was ordinal, and that it was converted to numeric data
via a mapping.
To the Provost the current professorial salary scale is simply a set
of facts that have, at this instant in time, certain fixed values
amongst a vast sea of possibilities.
In other words, the Provost sees the amounts being paid as an
accurate measurement in regard to the attribute ``professorial salary
payments''.

Nor is the Provost concerned by the fact that the average salary
value cannot be inverse-mapped to a professorial rank (indeed,
perhaps the Provost is a demographer, and hence equally comfortable
with the fact that the average couple have $1.93$ offspring).
If the Provost did want to compute the inverse mapping, the best that
can be said is that the average salary-weighted professorial rank
lies in the interval $[{\Lab{P2}}, {\Lab{P3}}]$; that is, exactly the
same as can be said for the median professorial rank once the
promotion has taken place.

It is also perfectly appropriate to apply a {\emph{categorical to
numeric}} mapping to categorical datasets.
The university might have an agreed mapping that, for each of the
class labels listed in Table~\ref{tbl-categorical}(b), specifies the
workload fraction available for teaching duties:
\begin{center}
\begin{tabular}{l cccc}
\toprule

Class     
	& \Lab{AM} & \Lab{RO} & \Lab{TO} & \Lab{TR}
\\
Teaching fraction
	& $0.20$ & $0.10$ & $0.80$ & $0.35$
\\
\bottomrule
\end{tabular}
\end{center}
The dataset of ten ``academic work type'' categories associated with
the same ten professors could thus be mapped to a dataset of ten
numeric ``teaching fractions''; and then those ten fractions could be
summed and averaged to determine, respectively, the total teaching
capacity of the university, and the average teaching fraction per
employed professor.

\myparagraph{Interval Scales}

The third level in Stevens' hierarchy corresponds to numeric data for
which the difference between any pair of values has meaning, but the
values themselves do not necessarily have direct interpretation (or
may, but it is somewhat arbitrary).
As an example, consider the timestamps employed in the Unix operating
system, which are measured in seconds since 1 January 1970, UTC.
At the time this sentence was being planned, the ``time'' was
indicated by {\Lab{1636099886}}; now that the sentence has been
(nearly) typed, the measurement is {\Lab{1636100081}}.\footnote{Both
obtained from {\url{https://www.unixtimestamp.com/index.php}}.}
Each of those two large numbers is, in isolation, somewhat
meaningless; but the difference between them has a clear
interpretation -- the two time measurements were $195$ seconds apart.
Planning and then composing that one sentence took over three
minutes.

Another example is given by the Celsius and Fahrenheit temperature
scales.
According to one, water freezes at {\degrees{0}}; according to the
other, at {\degrees{32}}.
Nevertheless, the two scales measure the same underlying attribute:
each one degree rise in temperature corresponds to the addition of a
fixed amount of thermal energy to a specified volume of water.
That fact holds regardless of whether the one degree increase is
between $\degrees{40}$ and $\degrees{41}$, or between $\degrees{73}$
and $\degrees{74}$, provided that either Celsius or Fahrenheit is
used for both components of the comparison.

On the other hand when time is measured in ``years'', it is not an
interval scale measurement relative to the underlying attribute of
``orbits of the sun''.
It is a good approximation, and most of us would be willing to say
that {2021} is ``tens years after'' {2011} in the same manner as
{2011} is ``ten years after'' {2001}; and are also willing to accept
that the cultural basis for selecting the reference year -- the
beginning of the current monarch's rule; or the birth or death of
some historical religious figure -- is arbitrary.
But the span from {2001} to {2010} inclusive contains $3652$ days and
$315{,}532{,}800$ seconds, whereas the span from {2011} to {2020}
contains $3653$ days and $315{,}619{,}200$ seconds.
More importantly, the span from {2001} to {2010} inclusive contains
$9.9988$ solar orbits, whereas the period from {2011} to {2020}
contains $10.0016$ solar orbits.
That is, the interpretation attached to intervals measured in years
as a surrogate for ``solar orbits'' differs according to whereabouts
in the scale those intervals are taken.
Nor are days or seconds linearly translatable into years.\footnote{Strictly speaking, nor is Unix time an interval scale,
because international time-keepers insert the occasional ``leap
second'' too, most recently making 31 December 2016 one second longer
than the $86{,}400=24\times60\times60$ seconds of a standard day, see
{\url{https://en.wikipedia.org/wiki/Leap_second}}.
Unix timestamps assume that there are always exactly $86{,}400$
seconds every day; and at any leap-second boundaries, that extra
second is achieved by subtracting one from the operating system's
internal time variable, and observing the same second a second time.}

The measurement points of an interval scale that are used in any
particular set of measurements or observations are {\emph{not}}
required to be equi-distant on the number line.
With the exception already noted, Unix ``seconds'' are always one
second apart, and ``days'' are normally one ``rotation of the earth''
apart.
But ``first day of the month'' dates when expressed as Unix
timestamps are not at fixed intervals (at least, not in the current
Gregorian calendar); similarly, ``business days'' is a valid
interval-based measurement that bypasses two days every seven.
In general, measurement points and measured values can be as close to
or far apart from each other as is consistent with accurate
representation of the underlying attribute that is being recorded and
in accordance with the purpose for which it is being measured.

The same flexibility extends to categoric to numeric mappings, and to
ordinal to numeric mappings.
It is perfectly acceptable for the salary increments between
professorial ranks to be of different sizes in
Table~\ref{tbl-ordinal}(b); and for the ordinal to numeric mapping
$\Num{\cdot}$ shown in Table~\ref{tbl-ordinal}(c) to make the
interval between {\Lab{WA}} and {\Lab{WR}} twice as large as the
interval between {\Lab{WR}} and {\Lab{R}}.
In the first case the decision on target values would have been made
by the university as a reflection of the cost of attracting and
retaining staff of the required caliber; in the second, the decision
on those relative intervals would have been made by the PC Chairs for
the conference in question, based on their experience of referee
behavior and the outcomes they sought via the paper review process.
Once established, any such mapping allows the class labels to be
converted to numbers, and for differences to be computed and
compared.
The fact that in Table~\ref{tbl-ordinal}(c) the target value $0$ is
not generated by any of the six label options is of no concern;
$-0.5$ and $+5$ are not amongst the mapped targets either.
Nor is $\$925$ a salary level that is available in
Table~\ref{tbl-ordinal}(b).

Datasets based on interval scales allow translation operators
($mx+c$, where $m>0$ and $c$ are constants) to be applied without
affecting relativities, and, as noted, for differences between
measurements, and ratios of differences between measurements, to be
compared; but not ratios of measurements themselves.
Consider the ordinal to numeric payroll mapping shown in
Table~\ref{tbl-ordinal}(b).
That mapping means that it is valid to both compute differences and
also to attribute meaning to the ratios between differences.
For example, $\Num{\Lab{P4}}-\Num{\Lab{P3}} = 1.5 \times
(\Num{\Lab{P3}}-\Num{\Lab{P2}})$, and it is evident that promotion to
{\Lab{P4}} from {\Lab{P3}} results in a pay-rise that is $1.5$ times
larger than the pay-rise that our friend received earlier when they
were awarded their promotion to {\Lab{P3}}.

Ratios between intervals defined by one scale might not correspond to
comparable intervals on a different scale that represents a different
underlying attribute.
For example, the {\Lab{P1}} professor who gets promoted to {\Lab{P2}}
might gain the same added {\emph{utility}} from their modest $\$50$
pay-rise as a {\Lab{P2}} professor gains when promoted to {\Lab{P3}}
and receives a $\$100$ increase in their weekly pay.
If we wish to map professorial classes to perceived utility of income
we are measuring a different underlying attribute, and should use a
different ordinal to numeric mapping.
On the other hand, Celsius and Fahrenheit do measure the same
underlying attribute, and one scale is thus a translation of the
other.

When the measurements are made on a continuous scale
the values in the dataset might have varying degrees of precision.
We can count weekly salaries down to the cent or even sub-cent level
if we wish to, or stick to whole dollars, or have a mixture.
Similarly, temperatures might be expressed as integers sometimes, or
to three decimal places at others; and a landscape gardener planning
a paling fence measures their $50$ meters to less precision than does
the builder of the swimming pool for an upcoming Olympics.
This is not an issue.
The requirement for an interval scale is {\emph{solely}} that taking
differences must always yield values that can be compared to each
other as ratios and hence be assigned meaning relative to the
underlying attribute; and that those interpretations must be
invariant with respect to {\emph{where}} in the scale the differences
arise.

Monetary amounts, distances, weights, and so on, all result
in datasets that have interval scale properties.
A $40$ kilogram weight is heavier than a $30$ kilogram weight by
exactly the same $10$ kilogram difference as a $25$ kilogram weight
is heavier than a $15$ kilogram weight.
And the last kilometer of a cycling race is exactly the same length
as the first kilometer, regardless of how long the race is.
That final kilometer might require more mental resilience than the
first, and it might require more muscle energy production too -- both
of which are underlying attributes that are not ``distance'', and
hence cannot be measured in units of kilometers -- but it will
certainly be one kilometer long.

When a dataset is presented on an interval scale (or following the
process of mapping a categorical or ordinal dataset to obtain a
derived interval-scale dataset), all of the operations permissible on
ordinal-scale measurements are again permissible.
In addition, a cumulative frequency distribution can be plotted as a
continuous (perhaps stepped) line with the horizontal axis determined
by values (rather than as a sequence of labeled bars in an
equi-spaced bar chart); and it is also permissible to compute the
arithmetic mean (average).
As a geometric interpretation, the arithmetic mean is the point
$\bar{p}$ at which the sum of the signed differences $p_i-\bar{p}$
for the elements $p_i$ in the dataset is zero, confirming that the
relationship between the arithmetic mean and the elements in the
dataset is invariant to the possible arbitrariness of the origin
point and multiplicative scale of the measurements.

\myparagraph{Ratio Scales}

When data is measured using a ratio scale, the data elements
themselves have meaning, as well as their differences; and the ratio
between data elements is a permissible computation that has the same
interpretation across the measurement scale.
For example, weight measured in kilograms is a ratio scale, with $20$
kilograms twice as heavy as $10$ kilograms in the same way that $50$
kilograms is twice as heavy as $25$ kilograms, and in the same way
that $44.092$ pounds (that is, $20$ kilograms) is twice as heavy as
$22.046$ pounds.
Consistency of ratios means that the zero point of the scale is no
longer arbitrary, and that it must be in a single unique location for
all ways of measuring that underlying attribute.

All of weight (in kilograms), distance (in kilometers), money paid as
salary (in dollars), and temperature (in degrees Kelvin, but not in
degrees Celsius of Fahrenheit) are ratio scales.
Plus, if for some reason we are specifically interested in time since
1 January 1970, then Unix timestamps are a ratio scale, with
{\Lab{50000000}} being twice as distant from 1 January 1970 as is
{\Lab{25000000}}.
On the other hand, if we have no reason to attach significance to 1
January 1970, then Unix timestamps are (only) an interval scale.
Similarly, the referee score mapping shown in
Table~\ref{tbl-ordinal}(c) is not a ratio scale.

\myparagraph{Absolute Measurements}

If the attribute that is being measured is one that can be directly
quantified, then that value can be used as the measurement without
further transformation.
For example, ``number of children'' is an absolute attribute (taking
on values zero, one, two, and so on) that does not require that
``units'' be specified -- compare with length measured in centimeters
or inches or yards or meters (or light-years or parsecs).
This category is not included in Stevens' taxonomy, but for
completeness it makes sense to note it here.

\myparagraph{Interpretable Outcomes}

If the observer has complete freedom to choose an ordinal to numeric
mapping, then little interpretation can be placed on any computed
attributes, such as the mean.
We could get a different outcome by choosing a different mapping.
But if the ordinal to numeric mapping is defined by the context in
which the dataset was created, and is bound to a set of target values
by some external reality -- as was the case, for example, with the
professors' salaries -- then that factual relationship makes the
mapping's values meaningful, and hence interpretable in terms of the
attribute from which the measurement was derived.
In the main example of this section, the Provost knew the total cost
of the ten professors because their salaries were defined via an
agreed and published mapping that could be summed, a real-world
consequence of the professorial ranks.

Values that are derived via some mapping can only be interpreted
{\emph{in the context of that mapping}}, and if the mapping changes,
so too will the derived values, and perhaps even the relativities.
If the ten professors decamp and move en masse to another university
(while retaining their current ranks), they are likely to be subject
to a different set of salaries.
If so, a different ordinal to numerical mapping will apply, and after
calculating their average salary relative to that university's pay
scales, their new Provost might reach a different conclusion about
the average salary-weighted professorial rank.

Similarly, if the conference PC chairs used a different mapping from
referee acceptance grades to numbers then the submitted papers might
get sorted into a different overall ``average paper score'' ordering,
and a different set of papers might be accepted.
But provided the mapping is defined by the PC chairs prior to the
outcomes of using it being examined, and is based on principles that
they believe can be successfully argued, then calculating average
referee scores is defensible, even though the relationship between
the mean of a mapping-derived dataset and the mapping's set of target
values is not required to be invariant to mapping changes.
This cause-and-effect nexus between mapping and conclusions is
both normal and acceptable, and provided the mapping has its basis in
the real world attribute that is the focus of the measurement, should
not be regarded as being proscribed in any way.

\myparagraph{Numbers Don't Remember}

In a parable involving ``football numbers'', {\citet{lord53}}
observes (giving an opinion via the voice of the ``statistician'' in
the story) that: ``{\emph{The numbers don't know that, \dots Since
the numbers don't remember where they came from, they always behave
just the same way, regardless}}''.
This statement has provoked a great number of words of commentary,
both in support and in opposition; with {\citet{sb09mathpsych}}
giving one of the more recent -- and also more insightful -- analyses.

The complementary argument made in this section is that if you
{\emph{do}} know where the numbers came from and why they have the
values that they do, and are confident that those values can be
justified in reference to the real world attribute that the mapping
is designed to represent, then those numbers may be used in your
analysis and interpretation of that real world equivalence.
The next section applies that principle to effectiveness measurement
in information retrieval.
 \section{Measurement in IR}
\label{sec-serps}

\myparagraph{Search Engine Result Pages}

Evaluation in batch (offline) information retrieval centers on
{\emph{search engine result pages}}, or SERPs, see
{\citet{sanderson10fntir}} for an overview.
A SERP is an ordered permutation of the $n$ documents in the
collection managed by the search service, or a $k$-element prefix of
such a permutation; and is the visible output that is presented to a
user in response to a query.
Each item in the SERP is a either a document, or a surrogate summary
of a document referred to as a {\emph{snippet}} or {\emph{caption}}.
In most IR batch evaluation methodologies (but not all) users who
examine a snippet in a SERP are regarded as having also examined the
document behind the snippet, and we will continue with that
assumption here.

The primary underlying attribute in IR evaluation is the
{\emph{usefulness}} of a SERP in terms of how well it addresses the
information need that provoked the user's query.
To that end, ``usefulness'' can be defined either as a combination of
{\emph{correctness}}, {\emph{coverage}}, {\emph{comprehensivity}} and
{\emph{cost of consumption}} of the information conveyed by the
SERP's documents, or in terms of the user's {\emph{satisfaction}}
after they have consumed the SERP.
Comparative evaluations based on usefulness are then used to
determine which search service, or parameter settings within a single
service, gives rise to the SERPs with the greatest
usefulness.\footnote{Strictly speaking, in a commercial environment
the economic objective is to maximize the expected total of all
future revenue derived from the universe of users (current and
potential) when discounted into today's dollars, subject to an
approved risk profile.
The ``all future'' component of that responsibility is why companies
invest in research initiatives and new product development, and also
why they must retain current customers by providing an attractive and
useful service.
And the ``potential users'' component is why social equity,
diversity, responsible sourcing, environmental sustainability, and so
on, are also important.}

SERPs are categorical data, because they are ordered $n$-tuples (or
ordered $k$-tuples) of documents, which are themselves categorical.
A standard assumption is that the individual documents making up the
SERP can each be assigned a per-document value known as
{\emph{relevance}}, corresponding to their in-isolation usefulness in
response to the query.
In most experimental contexts document relevance is represented on an
ordinal scale, based on {\emph{relevance grades}}.
The simplest possible measurement scale is a binary one, with labels
``\Lab{G0}'' meaning ``non-relevant'' and ``\Lab{G1}'' meaning
``relevant''.
Graded relevance scales make use of more classes, see, for example,
the one already shown in Table~\ref{tbl-ordinal}(d).
Relevance scales based on arbitrary numeric values are also possible,
and do not alter the arguments presented here.

Given that each document in a SERP can be assigned a relevance grade,
SERPs can be assigned to categorical classes based on their ordered
sequences of $n$ (or $k<n$) relevance grades.
For example, a five item SERP provided in response to a query might
be a member of the class
$\langle\Lab{G1},\Lab{G3},\Lab{G0},\Lab{G0},\Lab{G2}\rangle$,
with the five ordinal document relevance classes as defined in
Table~\ref{tbl-ordinal}(d).

\myparagraph{Counting and Ordering SERPs}

Even in the simplest case, with binary document relevance classes,
the number of SERP classes is huge.
If $n_0$ is the number of {\Lab{G0}} non-relevant labels across the
$n$ documents, and $n_1$ is the number of {\Lab{G1}} relevant labels,
then there are a total of $n!/(n_0!n_1!)$ different SERP classes.
Even when a $k$-element prefix of the SERP is taken, with
$k\le\min\{n_0,n_1\}$, there are still $2^k$ different SERPs.

It was noted above that SERPs are categorical data; nevertheless,
some SERP relativities can be derived from the ordering embedded in
the document relevance scale.
For example, when the five-element SERP
$\langle\Lab{G1},\Lab{G3},\Lab{G0},\Lab{G0},\Lab{G2}\rangle$,
is compared with the SERP
$\langle\Lab{G1},\Lab{G2},\Lab{G0},\Lab{G0},\Lab{G1}\rangle$,
it is apparent that the second one cannot possess more usefulness
than the first one, as it is less relevant in two document positions,
and equal in the other three dimensions.
More generally, we can be confident that SERP {\Lab{S1}} is
{\emph{non-inferior}} to SERP {\Lab{S2}} (denoted
$\Lab{S1}\succeq\Lab{S2}$) by considering two monotonicity
relationships:
\begin{description}[leftmargin=7mm,rightmargin=3mm]
\item{\emph{Rule 1}:}
SERP {\Lab{S1}} is non-inferior to SERP {\Lab{S2}} if every element
of {\Lab{S1}} is greater than or equal to the corresponding element
of {\Lab{S2}} in terms of their ordinal document relevance labels;
\item{\emph{Rule 2}:}
SERP {\Lab{S1}} is non-inferior to SERP {\Lab{S2}}
if {\Lab{S2}} can be formed as a
transformation of {\Lab{S1}} in which one or more elements are
swapped rightwards and exchanged with elements of strictly lower
document relevance that move leftward.
\end{description}
Rule~1 is an absolute relationship that does not rely on the
documents in the SERP being examined in a top-down manner
(left-to-right in the examples employed here).
Rule~2 arises from adding the assumption that the SERP is examined
sequentially from left-to-right, but is still not equivalent to
lexicographic ordering.

\begin{figure}[t]
\centering
\includegraphics[width=90mm,page=1,clip=true,trim=57mm 85mm 43mm 5mm]{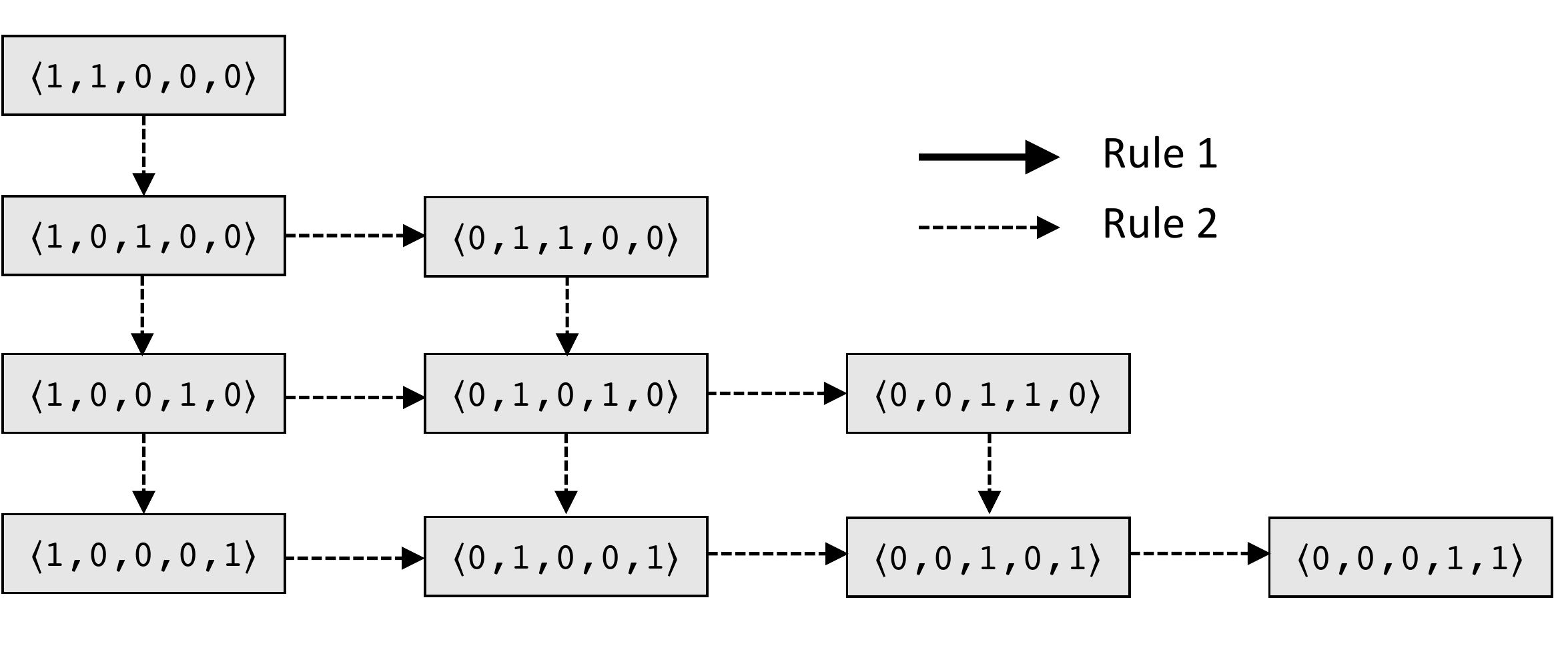}
\\(a) When $n=5$, $n_0=3$, and $n_1=2$
\\
\includegraphics[width=90mm,page=2,clip=true,trim=57mm 85mm 43mm 0mm]{diagrams.pdf}
\\(b) When $n=5$, $n_0=3$, $n_1=2$, and $k=4$
\mycaption{Hasse diagrams showing: (a) all SERPs of $n=5$ documents
in which there are $n_0=3$ non-relevant and $n_1=2$ relevant
documents; and (b) the set of $k=4$ prefixes of those SERPs.
The solid arrows indicate the ``Rule 1'' $\succeq$ relationships, and
the dotted arrows show ``Rule 2'' $\succeq$ relationships.
Two relevance grades are assumed, with ${\Lab{1}} > {\Lab{0}}$.
\label{fig-hasse-n5}}
\end{figure}

Figure~\ref{fig-hasse-n5}(a) shows the set of $\succeq$ relationships
when SERPs of $n=5$ binary document relevance grades with $n_0=3$ and
$n_1=2$ are considered, with ``{\sf{0}}'' and ``{\sf{1}}'' used as
shorthand for the document relevance labels {\Lab{G0}} and {\Lab{G1}}
respectively.
No Rule~1 relationships are possible when all $n$ documents are
included, because the SERPs must be permutations of each other.
Figure~\ref{fig-hasse-n5}(b) then shows the non-inferiorities that
arise when binary SERPs over $n=5$ documents (still with $n_0=3$ and
$n_1=2$) are truncated at $k=4$ for the purposes of evaluation.
Now Rule~1 relativities also occur.

All of the relationships shown in Figure~\ref{fig-hasse-n5} are
transitive, meaning that SERP pairs that are not linked by a directed
path of arrows are incomparable.
In both the $n=5$ case and the $k=4$ version the two axiomatic rules
are insufficient to impose a preference ordering between
$\langle1,0,0,1,0\rangle$ and $\langle0,1,1,0,0\rangle$.
That is, in the absence of any further information in regard to what
it is that users find to be useful, either of these two SERPs might
be preferred.

\myparagraph{IR Effectiveness Metrics}

Given that context, an {\emph{effectiveness metric}} is a categorical
to numeric mapping that assigns a real-valued number to each possible
class of SERP.
Those derived values are often, but not always, in the range
$0\ldots1$.
For example, assuming binary document relevance categories with class
labels {\Lab{G0}} and {\Lab{G1}}, the simple metric ``precision at
$k$'' (Prec@$k$) is computed as the number of relevant (\Lab{G1})
documents among the first $k$ in the SERP, divided by $k$.
Both $\langle1,0,0,1,0\rangle$ and $\langle0,1,1,0,0\rangle$ have
Prec@$5$ scores of $0.4$, and both have Prec@$4$ scores of $0.5$.
But $\langle0,1,1,0,0\rangle$ is deemed to be better than
$\langle1,0,0,1,0\rangle$ according to Prec@$3$.
Moreover, there is a real-world interpretation of Prec@$k$ that can
justify its use as a metric: if we suppose that each user of the
search system examines exactly the first $k$ documents in each SERP
they receive, then Prec@$k$ measures the fraction of documents viewed
by the user that are relevant.
That is, there is a clear connection between Prec@$k$ and an aspect
of the real-world situation that can be argued as being a way of
assessing SERP usefulness.

A wide range of other effectiveness metrics have been proposed to
augment Prec@$k$, including top-weighted ones that allocate
decreasing importance to documents the further they are from the head
of the SERP.
For example, the metric RR is defined (again, for binary relevance
grades) as the reciprocal of the index of the first position in the
SERP that contains a {\Lab{G1}} document.
The same two SERPs $\langle1,0,0,1,0\rangle$ and
$\langle0,1,1,0,0\rangle$ thus have RR values of $1.0$ and $0.5$
respectively.

With these definitions, an RR value of $0.25$ is possible, but an RR
value of $0.75$ is not.
Similarly, a Prec@$3$ value of $0.5$ not possible, nor a Prec@$5$
value of $0.35$.
Should we be concerned by these absences?
{\citet{fuhr17forum}} and the sequence of papers by
{\citet{ffp19ieeekade,ffl20irj,fff21ieeeaccess}} argue that in the
case of RR we most definitely should.
In particular, the first of {\citeauthor{fuhr17forum}}'s list of ten
``common mistakes'' (even if not outright commandments) is
``{\emph{Thou shalt not compute RR}}'', a directive justified
by ``{\emph{\dots the difference [in score] between ranks $1$ and $2$
is the same as that between ranks $2$ and $\infty$.
This means that RR is not an interval scale, it is only an ordinal
scale.}}''
We disagree with that conclusion.

\myparagraph{Salaries for SERPs}

To set the professorial salary levels listed in
Table~\ref{tbl-ordinal}(b) the university in question may have
engaged remuneration consultants and asked them to undertake a
comparative study of current real-world salary expectations for
professors of certain specified abilities.
The university knows it has to offer competitive salaries if it is to
retain staff, but under the ever-watchful eye of the Provost, doesn't
want to pay too much.
That is, we can assume that the correspondences listed in
Table~\ref{tbl-ordinal}(b) have been determined to be ``market
rates'' in some way, and drawn from a larger set of initial
possibilities that were considered as options.
The intervals between the salary points are meaningful; they
represent salary differentials that must be paid in a competitive
market, measured in dollars.

Suppose that a search engine company -- ``{\company}'', perhaps --
undertakes a similar market rates study.
They run surveys, host focus groups, meet with psychologists, and
sponsor IR-related conferences; and conclude from their
investigations that the great majority of {\company} users fall into
a ``shallow-hasty-youthful'' demographic that is highly focused on
getting a single correct result for each of their queries.
The study participants were also asked to estimate the monetary value
of example SERPs, and out of an immense amount of data a set of
correspondences between SERP classes (that is, SERP categories
constructed using binary document relevance grades {\Lab{G0}} and
{\Lab{G1}}, as shown in Figure~\ref{fig-hasse-n5}) and perceived
values emerges:
\begin{center}
\begin{tabular}{l ccccc}
\toprule
SERP type
	& \Lab{T1} & \Lab{T2} & \Lab{T3} & \Lab{T4} & $\cdots$
\\
Value      & $1.00$c & $0.50$c & $0.33$c & $0.25$c & $\cdots$
\\
\bottomrule
\end{tabular}
\end{center}
where the group {\Lab{T1}} contains all SERPs that commence with a
fully relevant document, $\langle\Lab{G1},\ldots\rangle$; group
{\Lab{T2}} contains all SERPs that have their first relevant document
in the second position and commence with
$\langle\Lab{G0},\Lab{G1},\ldots\rangle$; group {\Lab{T3}} contains
all SERPs that commence with two non-relevant documents and then have
a {\Lab{G1}}-grade document in third position,
$\langle\Lab{G0},\Lab{G0},\Lab{G1},\ldots\rangle$; and so on.
These grouped SERP categories -- {\Lab{T1}}, {\Lab{T2}} and so on --
form an ordinal arrangement, because of the positional references to
``first'' and ``second'', but the groups can also still be thought of
as categorical labels.

The company's chief financial officer (CFO) takes great interest in
this data.
To estimate the possible income should {\company} move to a user-pays
income model, the CFO further assembles a sample of ten recent
queries and the SERPs that were returned for them, and constructs
this dataset of SERP groups:
\[
\{  \Lab{T1},\Lab{T3},\Lab{T1},\Lab{T4},\Lab{T3},
            \Lab{T1},\Lab{T1},\Lab{T1},\Lab{T2},\Lab{T3} \}\,
	\, .
\]
The mode of this dataset is {\Lab{T1}}; in an ordinal sense the
median is either {\Lab{T1}} or {\Lab{T2}}; and it is meaningless to
ask about the ``average SERP category''.
But the {\company} CFO continues, joining the per-SERP revenues
estimates from the market research to the sample SERP distribution:
\begin{center}
\begin{tabular}{l cccc}
\toprule
SERP group
	& \Lab{T1} & \Lab{T2} & \Lab{T3} & \Lab{T4}
\\
Count
	& $5$ & $1$ & $3$ & $1$
\\
Revenue each
	& 1.00c & 0.50c & 0.33c & 0.25c
\\
Income
	& 5.00c & 0.50c & 1.00c & 0.25c
\\
\bottomrule
\end{tabular}
\end{center}
Summing the bottom row tells them that their current revenue
expectation from a user-pays model would be $6.75$c for this sample
of ten queries, or $0.675$c per query on average.

Now the critical question arises: is the computation of the average
payment per query using this framework a valid computation?
We argue that it is, and that it is meaningful in exactly the same
way that the average professorial salary is a meaningful value.
Both the professorial salaries and the per-SERP payments are on the
interval scale of ``money'', and by design (and expenditure on
consultant fees) reflect their respective real-world situations.
Hence, ``mean value per SERP'' is a valid measurement of search
according to its underlying attribute -- the usefulness of SERPs to
users, provided only that we are willing to equate the attribute of
usefulness and the attribute of value.
But that equivalence is one of the underpinning assumptions of
economics: that the price that someone is willing to pay for goods or
a service reflects the utility (that is, usefulness) that they expect
to derive from it.

\myparagraph{Different User Behaviors}

\begin{table}[t]
\centering
\mycaption{All possible SERPs composed of $n=5$ binary document
relevance grades containing $n_1=2$ relevant and $n_0=3$ non-relevant
documents.
The metric Prec@$5$ is $0.4$ for all ten SERPs.
There are no violations of the $\succeq$ relationships captured by
the arrows in Figure~\ref{fig-hasse-n5}.
\label{tbl-example}}
\newcommand{\tabent}[1]{\makebox[12mm]{#1}}
\renewcommand{\tabcolsep}{2.5mm}
\sisetup{
group-separator = {,},
round-mode = places,
round-precision = 3,
table-format = 1.3,
}\begin{tabular}{c c SSSS c S}
\toprule
\multirow{2}{*}{SERP}
	&& \multicolumn{4}{c}{Evaluated at $n=5$}
		&& \multicolumn{1}{c}{$k=4$}
\\
\cmidrule{3-6}\cmidrule{8-8}
	&& {\tabent{RR}}
		& {\tabent{RBP${}_{0.5}$}}
			& {\tabent{AP}}
				& {\tabent{NDCG}}
					&& {\tabent{Prec@4}}
\\
\midrule
$\langle1,1,0,0,0\rangle$
	&& 1.0000
		& 0.7500
			& 1.0000
				& 1.0000
					&& 0.5000
\\
$\langle1,0,1,0,0\rangle$
	&& 1.0000
		& 0.6250
			& 0.8333
				& 0.9197
					&& 0.5000
\\
$\langle1,0,0,1,0\rangle$
	&& 1.0000
		& 0.5625
			& 0.7500
				& 0.8772
					&& 0.5000
\\
$\langle1,0,0,0,1\rangle$
	&& 1.0000
		& 0.5313
			& 0.7000
				& 0.8503
					&& 0.2500
\\
$\langle0,1,1,0,0\rangle$
	&& 0.5000
		& 0.3750
			& 0.5833
				& 0.6934
					&& 0.5000
\\
$\langle0,1,0,1,0\rangle$
	&& 0.5000
		& 0.3125
			& 0.5000
				& 0.6509
					&& 0.5000
\\
$\langle0,1,0,0,1\rangle$
	&& 0.5000
		& 0.2813
			& 0.4500
				& 0.6241
					&& 0.2500
\\
$\langle0,0,1,1,0\rangle$
	&& 0.3333
		& 0.1875
			& 0.4167
				& 0.5706
					&& 0.5000
\\
$\langle0,0,1,0,1\rangle$
	&& 0.3333
		& 0.1563
			& 0.3667
				& 0.5438
					&& 0.2500
\\
$\langle0,0,0,1,1\rangle$
	&& 0.2500
		& 0.0938
			& 0.3250
				& 0.5013
					&& 0.2500
\\
\bottomrule
\end{tabular}
 \end{table}

The reader will doubtless have noted that the SERP ``pricing''
mechanism used in that previous example corresponds to Reciprocal
Rank, RR.
What if {\company}'s market research also noted other factors that
influence the amount that a user is willing to pay, in addition to
the position of the first relevant result in the SERP?
For example, suppose that an even more detailed user evaluation (and,
who knows, perhaps a deep convoluted neural model as well) reveals
that customers are willing to pay $0.50$c if the first document in
each SERP is relevant; plus (independently) $0.25$c if the second is
relevant; plus another $0.125$c if the third is relevant; and so on;
adding up the payments right through the length of the SERP.
The ``RBP${}_{0.5}$'' column in Table~\ref{tbl-example} shows the ten
different per-SERP values that can arise when this computation is
applied to the set of ten $n=5$, $n_1=2$ SERPs shown earlier in
Figure~\ref{fig-hasse-n5}, and places those derived scores beside the
corresponding RR values.
This new mapping function yields the effectiveness metric
{\emph{rank-biased precision}} (RBP) with parameter $\phi=0.5$
{\citep{mz08acmtois}}.

Both RR and RBP can be applied to the CFO's dataset of ten SERPs,
with different average scores emerging.
Those two averages must not be compared to each other, because they
were computed using different mappings and hence different
assumptions about value.
They may well be correlated, but are not convertible.
Nevertheless, both the RR and RBP${}_{0.5}$ averages over a dataset
of categorical SERPs are valid computations in the context of the
numeric mappings that were employed when computing them.
What the CFO must do is decide which conversion mechanism best
captures the value of each SERP class to the members of their user
base, that is, which context they believe is the most realistic
assessment of value as a surrogate for the underlying attribute of
SERP usefulness.

\myparagraph{Users, Models, and Metrics}

There are many other possible effectiveness metrics, and more are
proposed each year.
Table~\ref{tbl-example} adds two further options to the three that
have already been mentioned: {\emph{average precision}}, AP
{\citep{buckley05,robertson08sigir}}; and {\emph{normalized
discounted cumulative gain}}, NDCG {\citep{jk02acmtois}}.
Each of the five metrics shown is a categorical to numeric mapping,
with the numeric targets representing perceived utility, expressed in
units of ``willing to pay this many cents for a SERP in this
category'', and in which ``cents'' is an imaginary currency that
nevertheless has a fixed multiplicative exchange rate that allows
conversion to Euros, to USD, to JPY, to RMB, and so on.
Just as inches can be converted to parsecs.

The final two -- AP and NDCG -- are computed somewhat differently to
the three already described.
They involve a ``normalization'' step that adjusts the score
(payment) associated with each SERP according the maximum amount of
relevance available across the collection (in the binary examples
used here, expressed by the value $n_1$), adding an implication that
users are willing to pay increased amounts if relevant documents are
relatively scarce, but equally implying that users are somehow aware
of the scarcity or not of relevant documents in regard to each query
they issue.
Note also that all of RBP, AP, and NDCG are top-weighted, meaning
that if they are evaluated across the whole collection (that is, on
full-length SERPs of length $n$ rather than at-$k$ truncated ones),
Rule~2, noted above, results in strict superiority ($\succ$), rather
than non-inferiority~($\succeq$).

More generally, most IR metrics have a corresponding {\emph{user
browsing model}}, which hypothesizes the way in which users interact
with each SERP, and the subconscious process they follow as they
consume SERPs and assess usefulness -- the attribute that we are
trying to measure.
Thus, one way in which IR effectiveness metrics have been studied is
via the development of user browsing models of increasing
sophistication {\citep{atc2018sigir,zllzxm17sigir,
cmzg09cikm,mbst17acmtois,carterette11sigir,wsdm20wm}}.
Each such model maps a categorical SERP to a numeric assessment of
that SERP's value on the real number line, usually between $0.0$ and
$1.0$ inclusive, in units of ``expected utility gained per document
inspected'', using the corresponding browsing model as a guide to the
manner in which the user consumes, and ends their inspection of, the
SERP.

\myparagraph{Equi-Spaced Intervals}

As was noted earlier, {\citet{fuhr17forum}} and
{\citet{ffp19ieeekade,ffl20irj,fff21ieeeaccess}} criticize RR because
of this pattern of scores:
\begin{center}
first relevant document at rank one, RR${}=1.0$\\
	\raisebox{2.5mm}{\rotatebox{270}{$\rightarrow$}}\quad\makebox[0mm][l]{$-0.5$}\\
first relevant document at rank two, RR${}=0.5$\\
	\raisebox{2.5mm}{\rotatebox{270}{$\rightarrow$}}\quad\makebox[0mm][l]{$-0.5$}\\
no relevant document in top $k$, RR${}=0.0$.\\
\end{center}
In doing so, they overlook the possibility of users {\emph{wanting}}
those two intervals to be of equal importance.
If users' perceptions of usefulness concur with the relationship
between those three classes of SERP, then RR {\emph{is}} an interval
scale, with intervals between observed values at different points on
the scale that {\emph{do}} have the same interpretation in terms of
the underlying attribute of usefulness.
Any argument that RR is an unsuitable mapping must be justified based
on rhetoric (or on data) about user perceptions of usefulness, rather
than on non-uniformity of intervals.

{\citet[page 136193, in connection with their
Figure~3]{fff21ieeeaccess}} extend that earlier claim, writing:
``{\emph{the real problem with IR evaluation measures is that their
scores are not equi-spaced and thus they cannot be interval
scales}}''.
This assertion leads them to a proposal that existing metrics be
{\emph{intervalized}}, by enumerating all possible metric values over
truncated SERPs of some defined length ($k=10$, or $k=20$ say, but
certainly not $k=100$, because of combinatorial growth issues) and
then mapping the ordering implied by those values to a
uniform-interval scale to get new versions of those metrics.

To understand the process of intervalization, consider the metric
NDCG, already illustrated in Table~\ref{tbl-example}.
If we assume that $n_0,n_1\ge 3$ then there are $2^k=8$ different
binary-grade SERPs possible of length $k=3$, with NDCG$@3$ scores
(sorted by score, to three decimal places) of
\[
\{
0.000,
0.235,
0.296,
0.469,
0.531,
0.704,
0.765,
1.000
\}\,.
\]
That set of eight irregularly-spaced NDCG$@3$ scores would be
intervalized to the range $[0,1]$ via the corresponding
uniformly-spaced set of eight target values (all multiples of $1/7$,
again represented to three decimal places)
\[
\{
0.000,
0.142,
0.285,
0.428,
0.571,
0.714,
0.857,
1.000
\}\,.
\]
The mapped uniform-interval values would then be used to compute
means and as a basis for comparing systems, and to undertake
statistical tests, as a derived variant of NDCG$@3$.
Similarly, for the metric NDCG$@10$, a set of $1024$ mapped NDCG
values would be generated, at uniform intervals of
$1/1023$.\footnote{Note that here we make use of the ``Microsoft''
version of NDCG, in which the discount at rank $d$ is $\log_2(1+d)$
for all $d\ge 1$, whereas the examples provided by
{\citet{fff21ieeeaccess}} use the original {\citet{jk02acmtois}}
parameterized discount in which ranks $d\le b$ have a discount of
$1.0$, and ranks $d\ge b$ have a discount of $\log_b d$.
In the {\citet{jk02acmtois}} implementation, there are $768$ distinct
NDCG$@10$ values possible, and hence the intervalized version of this
metric would use a uniform interval of $1/767$.}

We believe that intervalization should regarded with scepticism.
There is no requirement in Steven's typology {\citep{stevens46}} that
interval scales be restricted to uniform distances between the
available measurement points; the requirement is simply that the
ratio between pairs of intervals be indicative of the corresponding
difference in the underlying attribute.
Altering the categorical to numeric mapping used to assign score to
SERPs changes the relativities being measured, and thus affects the
outcome of any subsequent arithmetic.
This effect is especially notable for the metric RR.
If truncated rankings of length $k$ are used, mapping to an
equi-spaced scale yields:
\begin{center}
first relevant at rank one\\
	\raisebox{2.5mm}{\rotatebox{270}{$\rightarrow$}}\quad\makebox[0mm][l]{$-1/k$}\\
first relevant document at rank two\\
	\raisebox{2.5mm}{\rotatebox{270}{$\rightarrow$}}\quad\makebox[0mm][l]{$-1/k$}\\
first relevant at rank three\\
$\cdots$\\
first relevant document at rank $k-1$\\
	\raisebox{2.5mm}{\rotatebox{270}{$\rightarrow$}}\quad\makebox[0mm][l]{$-1/k$}\\
first relevant document at rank $k$\\
	\raisebox{2.5mm}{\rotatebox{270}{$\rightarrow$}}\quad\makebox[0mm][l]{$-1/k$}\\
no relevant document in the first $k$\\
\end{center}
which is logically equivalent to using the rank of the first relevant
document as the assessment of SERP usefulness -- let's call it the
metric R1, sometimes referred to as ``expected search length''
{\citep{cooper68adoc}}.
While that is a perfectly valid measure, it probably isn't a
plausible way of measuring the underlying attribute of SERP
usefulness.
Would a user of an IR system really perceive having the first
relevant document at rank $100$ rather than at rank $99$ as being the
same amount less useful as is having the first relevant document at
rank $2$ rather than at rank $1$?
Indeed, R1 is sufficiently obvious as a possible metric that if it
were reflective of user perceptions of usefulness, then it would have
been in common use in IR evaluation for the last several decades.
There has been a reason why R1 has not been used as a measure of SERP
quality -- because it doesn't adequately reflect SERP usefulness.

Finally, note that even RBP${}_{0.5}$ does not guarantee a
uniform-interval scale, adding further confusion to the situation.
{\citet{fuhr17forum}} and
{\citet{ffp19ieeekade,ffl20irj,fff21ieeeaccess}} suggest that
RBP${}_{0.5}$ is an example of a valid (by their requirements)
interval scale IR metric, because if the runs being evaluated are of
length $k$, then all multiples of $2^{-k}$ between $0$ and $1-2^{-k}$
can be generated as metric scores, and hence the measurement points
are at uniform intervals.
But that conclusion is only correct when $k\le \min\{n_0,n_1\}$.
When complete SERPs to depth $n$ are scored via RBP${}_{0.5}$, or
when truncated SERPs are scored and $n_1<k$ (a situation that is by
no means improbable), the available set of RBP${}_{0.5}$ values is
not uniform-interval -- as is illustrated in Table~\ref{tbl-example},
comparing the sequence $0.565$, $0.531$, and $0.375$, for example.
Indeed, when there is a single relevant document for some query (a
navigational query {\citep{broder02forum}}, with $n_1=1$),
RBP${}_{0.5}$ gives this pattern of scores:
\begin{center}
relevant at rank one, RBP${}_{0.5}=0.5$\\
	\raisebox{2.5mm}{\rotatebox{270}{$\rightarrow$}}\quad\makebox[0mm][l]{$-0.25$}\\
first relevant document at rank two, RBP${}_{0.5}=0.25$\\
	\raisebox{2.5mm}{\rotatebox{270}{$\rightarrow$}}\quad\makebox[0mm][l]{$-0.25$}\\
no relevant document in the first $k$, RBP${}_{0.5}=0$,\\
\end{center}
exactly matching (modulo a linear transformation, viz, multiplication
by two, a permissible operation) the RR intervals over the same three
$k$-truncated SERP classes objected to by {\citet{fuhr17forum}}.
 \section{Other Considerations}
\label{sec-other}

\myparagraph{Related Work}

Fuhr's exposition {\citep{fuhr17forum}} addressing experimental
protocols in IR has also been commented on by {\citet{sakai20forum}}
who, amongst other concerns, writes (with acknowledgment to input
from Stephen Robertson): ``{\emph{it is also not clear to me whether
RR really cannot be considered as an interval-scale measure}}'', and
specifically questions the RR example given by Fuhr (first relevant
document at rank one versus first relevant document at rank two
versus no relevant document at all) and asks why this cannot ever be
congruent with the user's perception of SERP usefulness, thereby
anticipating our own concerns.
{\citet{sakai20forum}} goes on to present agreement rates between
human assessors and effectiveness metrics in regard to SERP quality
that summarize the results of an experiment by {\citet{sz19sigir}}
that compared SERPs in a side by side manner and elicited preferences
as to overall usefulness (in this case, via the question ``Overall,
which SERP is more relevant to the query?'').
It is experiments such as these that will establish which
effectiveness metrics best correlate with user perceptions of SERP
usefulness for various search applications and different
sub-demographics of users {\citep{sz21acmtois}}.
Similarly, consideration of perceived user experience is what has
driven much of the recent development of effectiveness metrics -- see
{\citet{mbst17acmtois}}, {\citet{zllzxm17sigir}}, and
{\citet{atc2018sigir}}, for example.

There has also been followup commentary in regard to Stevens'
original paper {\citep{stevens46}} about scales of measurement.
The contribution by {\citet{lord53}} has already been noted; amongst
many others the evaluations by {\citet{ta84}} and {\citet{vw93}} also
help delineate some of the issues that have emerged when considering
scales of measurement and their implications.
{\citet{sb09mathpsych}} provide a careful assessment of the role of
Lord's claimed ``counter example'' in regard to Stevens' taxonomy of
measurement scales.

In addition to the studies already discussed in
Section~\ref{sec-serps}, a range of work has considered the
underpinning measurements involved in IR.
For example, {\citet{bm12ictir}} consider measurement scales and SERP
orderings, developing and extending axiomatic relationships akin to
the ``Rule 1'' and ``Rule 2'' given above; and {\citet{ffm15ictir}}
undertake a similar exploration.
In a related study, {\citet{moffat13airs}} considers effectiveness
metrics in terms of a suite of seven numeric properties that they
might possess.
Other work -- for example, {\citet{th01sigir}}, {\citet{ts06sigir}},
{\citet{spck10sigir}}, {\citet{bcwcst10sigir}},
{\citet{llm-etal18www}}, and {\citet{zhang20sigir}} -- has considered
the extent to which whole-of-SERP usefulness is adequately captured
by current effectiveness metrics.

\myparagraph{Use of Recall Base}

{\citet{fuhr17forum}} and
{\citet{ffp19ieeekade,ffl20irj,fff21ieeeaccess}} note the
difficulties created by the use in some metrics of what they term the
``recall base'', the number $\RB$ of relevant (assuming only binary
relevance grades) documents in the collection for the topic in
question, denoted in Section~\ref{sec-serps} as the quantity $n_1$.
From the point of view of {\citeauthor{fuhr17forum}} and
{\citeauthor{ffp19ieeekade}}, those difficulties arise because
normalization by $\RB$ means that the set of generable measurement
points for any query in a set of topics might not numerically align
with the available measurement points for other topics that have
different values for $\RB$.
But it is worth noting that the recall base affects the available
measurements even when $\RB$ is not a visible component of the
effectiveness metric.
To observe this, consider Table~\ref{tbl-example} again.
It lists the ten possible SERP classes for a collection of $n=5$
documents and for queries with $n_1=2$ relevant documents in the
collection, together with metric score according to five metrics.
If another topic for the same test collection has $n_1=1$, the
RBP$_{0.5}$ scores are limited to the set $\{0.500, 0.250, 0.125,
0.063, 0.031\}$, none of which align with the set of available
$n_1=2$ RBP$_{0.5}$ scores listed in Table~\ref{tbl-example}.

Moreover, even when truncated rankings are considered, with (say)
$k=2$ used to calculate the scores, a topic for which $n_1=1$ is
unable to deliver a RBP${}_{0.5}$ score of $0.75$.
If RBP${}_{0.5}$ is to be used across a collection of topics, and if
exactly the same set of measurement points must be available for
every topic, then the SERP truncation length $k$ must satisfy
$k\le\min_i n_{1,i}$, where $n_{1,i}$ is the $n_1$ recall base
associated with the $i$\,th topic.
This places a severe limitation on any experiments making use of that
collection.
(The same restriction also applies to $n_0$, but in most retrieval
environments $n_1$ is smaller than $n_0$ by several orders of
magnitude.)

Our contention in this work is that the measurement scale is always
the positive real number line, and hence that no question of
alignment (or not) of measurement points across sets of topics
arises.
On the other hand, there are other reasons to eschew metrics that
make use of the recall base, based on the desire for effectiveness
metrics to reflect plausible user behaviors, and the user's inability
to actually know the value $n_1$ as they consider the SERP
{\citep{mz08acmtois,zmp09forum,lmc16irj}}.

\myparagraph{Graded Relevance and Gain Mappings}

The discussion above focused on binary-level document relevance
labels, but the same points apply to multi-level labels of the kind
suggested in Table~\ref{tbl-ordinal}(d).
Multi-level evaluations normally make use of two mapping stages.
The first converts ordinal document relevance classes to numeric
{\emph{gains}} via a {\emph{gain mapping}} function $\Num{\cdot}$
that converts ordinal document relevance grades to gain values in
$0\ldots1$, as shown, for example, in Table~\ref{tbl-ordinal}(d).
The second mapping then takes an $n$- or $k$-vector of numeric gain
values, combines them in a way that discounts gains as ranks
increase, and generates a single numeric score.
The metrics {\emph{discounted cumulative gain}} (DCG) and
{\emph{normalized discounted cumulative gain}} (NDCG)
{\citep{jk02acmtois}} make quite deliberate use of real-valued
document gains, as do RBP {\citep{mz08acmtois}} and {\emph{expected
reciprocal rank}} (ERR) {\citep{cmzg09cikm}}, with the goal of
providing more nuanced effectiveness measurements, and hence the
ability to respond with more sensitivity to perceived differences in
SERP usefulness {\citep{sor02sigir}}.
Average precision can also be broadened to make use of graded
document relevance categories {\citep{rky10sigir,dp10sigir}}.

The complex inter-relationships between the range of gain mappings
that might be employed, and then the metric mapping itself, further
mean that metric scores will not (and as is our firm contention here,
need not) result in uniform-interval measurements.

Gain mappings are also measurements, of course, pertaining to the
usefulness of individual documents.
For example, the ordinal class labels listed in
Table~\ref{tbl-ordinal}(d) might be included in a handbook provided
to assessors as part of their training, along with detailed
descriptions and examples.
Document gain labels -- the $r_i$ values used to compute
effectiveness metrics -- can also be more directly measured.
For example, magnitude estimation techniques
{\citep{tsmm15sigir,mmst17acmtois}}, side-by-side preference
elicitation {\citep{cbcd08ecir,spck10sigir,adcs18ymt,avyc21arxiv}},
and ordinal scales in which the class labels are numbers
{\citep{redm18sigir}} can all be used to develop numeric document
gain labels.

\myparagraph{Statistical Tests}

An important component of IR evaluation is the use of statistical
tests (see, for example, {\citet{sac07cikm}}, {\citet{sakai16sigir}},
and {\citet{ulh19sigir}}).
The appropriateness of any particular test depends in part on the
distributional conditions required by that test, and it should be
noted that our argument here in regard to metric values being numbers
that can be averaged is most definitely {\emph{not}} an argument that
all metrics can be tested with any particular statistical test.
Thoughtful selection of a statistical test, and, if necessary,
verification of any required distributional conditions governing its
applicability, must always be a critical part of IR experimental
design.
On the other hand, choosing an effectiveness metric because it is
amenable to a particular statistical test represents ``the tail
wagging the dog'' (pun intended), and is not a course of action that
should be considered.
The metric must be chosen first, and only then can the statistical
test be selected.

Similarly, we have no concerns with Fuhr's seventh rule, covering the
need for multiple hypothesis adjustments, but note that in the case
of test collection reuse it cannot always be properly achieved.
{\citet{cmycl21sigir}} provide an overview of some of these issues,
and {\citet{sakai20forum}} has also voiced opinions in support of
Fuhr's comments in regard to statistical testing.

 \section{Conclusions}
\label{sec-conclusions}

We have discussed the role of interval scale measures in information
retrieval evaluation and, via a sequence of examples, presented our
view that {\emph{all}} IR effectiveness metrics can be considered to
be interval scale measurements, provided only that the mapping from
SERP categories to numeric scores has a real-world basis and can be
argued in some way as corresponding to the underlying usefulness of
each SERP, as experienced by the users of that IR system.
That is, while care needs to be exercised when choosing the metric
that best fits the user experience for any particular IR application
(for example, the ``shallow-hasty-youthful'' users that form the
{\company} demographic), once that match has been decided, the values
calculated by the effectiveness metric may be used as ``simple
numbers'', without regard to their origins in a categorical-scale
SERP dataset.

Metric choice is a critically important design decision in any IR
experiment, and different metrics might lead to different outcomes
from a planned experiment.
But the choice between metrics should determined by the projected
user behavior and their implicit evaluation of ``usefulness''
relative to their search task, and not because of the regularity or
otherwise of the gaps between adjacent numeric values generated over
the universe of categorical SERP classes, and nor as a consequence of
amenability or system separability associated with any particular
statistical test.

In addition, we have argued that the proposed ``intervalization'' of
current IR effectiveness metrics is neither required nor helpful.
If the raw metric value is indeed a defensible measurement of SERP
usefulness and corresponds to the user experience when they are
presented with a member of that SERP category, then
equi-intervalizing those measurements via a different categorical to
numeric mapping must of necessity distort and alter any findings that
arise, and thus risks masking what would otherwise be valid
conclusions.
And if the raw metric is not a defensible measurement of SERP
usefulness, then equi-intervalizing its scores seems unlikely to
improve the situation.

\paragraph{Acknowledgment}
Joel Mackenzie, Tetsuya Sakai, Falk Scholer, and Justin Zobel
provided useful input.
This work was supported under the Australian Research Council's
Discovery Projects funding scheme (project DP190101113).
Science proceeds via debate, and we are grateful to be part of a
community in which debate is not only possible, but actively
welcomed.

\paragraph{Disclosure}
The author has no other relevant financial or non-financial interests
to disclose.

\renewcommand{\bibsep}{5pt}
\bibliographystyle{abbrvnat}

\end{document}